\documentclass[lettersize,journal]{IEEEtran}
\usepackage{amsmath,amsfonts}
\usepackage[ruled]{algorithm2e}
\usepackage{array}
\usepackage{bm}
\usepackage[caption=false,font=normalsize,labelfont=sf,textfont=sf]{subfig}
\usepackage{textcomp}
\usepackage{stfloats}
\usepackage{url}
\usepackage{verbatim}
\usepackage{graphicx}
\usepackage{cite}
\usepackage{multirow}
\usepackage{graphicx}
\usepackage{lscape}
\usepackage[table,xcdraw]{xcolor}

\title{QS-TTS: Towards Semi-Supervised Text-to-Speech Synthesis via Vector-Quantized Self-Supervised \\ Speech Representation Learning}

\author{
    Haohan Guo, ~\IEEEmembership{Student Member, ~IEEE},
    Fenglong Xie, ~\IEEEmembership{Member, ~IEEE}, \\
    Jiawen Kang, ~\IEEEmembership{Student Member, ~IEEE},
    Yujia Xiao, ~\IEEEmembership{Student Member, ~IEEE}, \\
    Xixin Wu, ~\IEEEmembership{Member, ~IEEE}, 
    Helen Meng, ~\IEEEmembership{Fellow, ~IEEE}
    \thanks{Hanhan Guo and Helen Meng are with the Human-Computer Communications Laboratory (HCCL), Department of Systems Engineering and Engineering Management, Chinese University of Hong Kong, Hong Kong SAR, China (e-mail: hguo@se.cuhk.edu.hk; hmmeng@se.cuhk.edu.hk). Xixin Wu is with Stanley Ho Big Data Decision Analytics Research Centre, The Chinese University of Hong Kong, Hong Kong SAR, China (e-mail: xixinwu@cuhk.edu.hk). Fenglong Xie is with Xiaohongshu, Beijing, China (e-mail: fenglongxie@xiaohongshu.com).}
    \thanks{This research was supported by the Center for Perceptual and Interactive Intelligence (CPII) Ltd under the Innovation and Technology Commission's InnoHK scheme.}
}

\begin{document}

\maketitle

\begin{abstract}

This paper proposes a novel semi-supervised TTS framework, QS-TTS, to improve TTS quality with lower supervised data requirements via Vector-Quantized Self-Supervised Speech Representation Learning (VQ-S3RL) utilizing more unlabeled speech audio. This framework comprises two VQ-S3R learners: first, the principal learner aims to provide a generative Multi-Stage Multi-Codebook (MSMC) VQ-S3R via the MSMC-VQ-GAN combined with the contrastive S3RL, while decoding it back to the high-quality audio; then, the associate learner further abstracts the MSMC representation into a highly-compact VQ representation through a VQ-VAE. These two generative VQ-S3R learners provide profitable speech representations and pre-trained models for TTS, significantly improving synthesis quality with the lower requirement for supervised data. QS-TTS is evaluated comprehensively under various scenarios via subjective and objective tests in experiments. The results powerfully demonstrate the superior performance of QS-TTS, winning the highest MOS over supervised or semi-supervised baseline TTS approaches, especially in low-resource scenarios. Moreover, comparing various speech representations and transfer learning methods in TTS further validates the notable improvement of the proposed VQ-S3RL to TTS, showing the best audio quality and intelligibility metrics. The trend of slower decay in the synthesis quality of QS-TTS with decreasing supervised data further highlights its lower requirements for supervised data, indicating its great potential in low-resource scenarios.

\end{abstract}

\section{Introduction}

Text-to-Speech (TTS) synthesis is a technology aiming to convert text to speech signals with correct pronunciation, natural prosody, and high audio fidelity. It has been widely used in various intelligent products, e.g. Human-Computer Interaction (HCI) \cite{mohasi2006text, guo2021conversational}, Speech-to-Speech Translation (S2ST) \cite{liu2021incremental, sudoh2020simultaneous}, and Artificial Intelligence Generated Content (AIGC) \cite{zhang2023survey}. Meanwhile, with the popularization and penetration of AI technology into various fields, the capability of a TTS system on personalization and customization has also been paid more attention, to serve all people around the world better. But the high cost of creating a TTS dataset with sufficient high-quality speech audio and accurate transcripts hinders the development of TTS in this aspect. Hence, reducing the training requirement for supervised data of a high-quality TTS system is becoming more urgent.

\begin{figure}[htp]
    \centering
    \includegraphics[width=8cm]{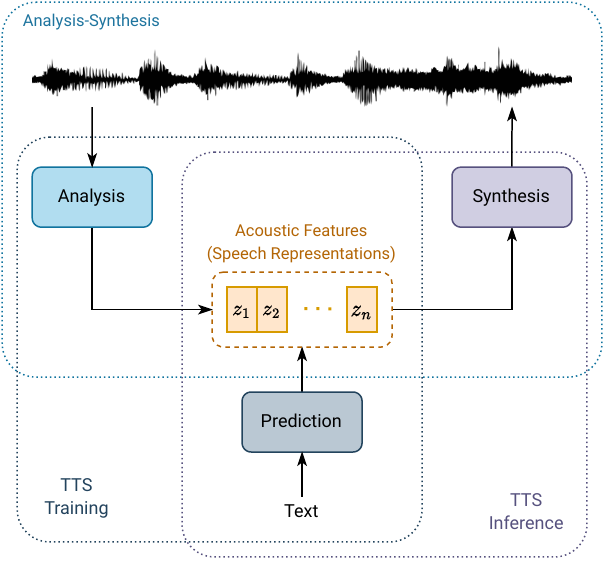}
    \caption{A mainstream TTS framework: the acoustic feature or speech representation is extracted from the waveform via the analysis module as the output of the prediction module, and decoded to the waveform by the synthesis module.}
    \label{fig:tts_framework}
\end{figure}

Fortunately, a TTS system does not rely on supervised data entirely. It is feasible to utilize more unlabeled speech data that is easier to collect to compensate for insufficient supervised data in TTS training. For example, in a mainstream modular TTS framework, as shown in Fig. \ref{fig:tts_framework}, we can employ unlabeled speech data to
\begin{itemize}
    \item Enhance the analysis module to provide practical data-derived speech representations with sufficient phonetic information and easier to predict from the text.
    \item Enhance the synthesis and prediction modules by pre-training them on relevant tasks.
\end{itemize}
In this way, we may reduce supervised data requirements by building a semi-supervised TTS system using more unlabeled speech data. And the key is seeking an appropriate approach or task utilizing unlabeled speech audio well to provide the expected features and pre-trained models to TTS.

In this regard, Self-Supervised Speech Learning (S3RL) has shown excellent capability in providing profitable speech representations and pre-trained models for various supervised speech tasks. Especially in speech synthesis \cite{polyak2021speech, guo2022msmc}, vector-quantization-based generative S3RL \cite{Oord2017NeuralDR} has already demonstrated its eminent performance, providing a compact, discrete speech representation to reduce modeling complexity while keeping high-quality speech reconstruction. Inspired by it, in this paper, we propose QS-TTS, a novel semi-supervised TTS framework based on Vector-\textbf{Q}uantized \textbf{S}elf-Supervised Speech Representation Learning (VQ-S3RL). 

This framework conducts VQ-S3RL on high-quality unlabeled speech data for two goals:
\begin{itemize}
    \item Providing profitable speech representations for TTS.
    \item Providing effective pre-trained models to enhance the acoustic model and the vocoder.
\end{itemize}
It comprises two learners. First, we train a principal VQ-S3R learner, which combines the contrastive S3RL model, HuBERT \cite{hsu2021hubert}, and the proposed generative S3RL model, Multi-Stage Multi-Codebook (MSMC) VQ-GAN in cascade. It converts the waveform into the MSMCR, a compact generative S3R comprising multiple sequences at different time resolutions and quantized by multiple codebooks, which is then decoded to high-quality speech audio via adversarial speech generation. Then, we train an associate VQ-S3R learner to abstract the MSMCR into a highly-compact VQ representation through a VQ-VAE-based model with multi-sequence encoding and decoding. These two generative VQ-S3R learners are applied in TTS training using supervised data, including playing the role of the analysis module providing the MSMCR, and providing pre-trained models for the prediction and synthesis module training.

In experiments, in addition to MOS tests to evaluate QS-TTS subjectively, we also measure the performance of QS-TTS objectively on audio quality using Frechét Distances (FD) in various embedding spaces and the intelligibility using the Character Error Rate (CER) and Phoneme Error Rate (CER). We first evaluate the overall performance of QS-TTS by comparing it with mainstream TTS approaches, e.g. FastSpeech, VITS, and their semi-supervised versions, in standard-resource and various low-resource scenarios. Then, we investigate the effect of the proposed VQ-S3RL on TTS by comparing it with different speech representations and transfer learning methods. Finally, we evaluate the performance of QS-TTS with different sizes of supervised data to further validate its effectiveness in reducing supervised data requirements.

In the rest of this paper, Section \ref{sec:bg} introduces the background of this work. Section \ref{sec:model} illustrates the framework of QS-TTS in detail. Experiments are described in Sections \ref{sec:exp} and \ref{sec:result}. Finally, Section \ref{sec:con} gives the conclusion to this paper.\footnote{Audio samples are available at \url{https://hhguo.github.io/DemoQSTTS/}}

\section{Background}
\label{sec:bg}

\subsection{Semi-Supervised Text-to-Speech Synthesis}

The semi-supervised TTS aims to utilize both supervised and unsupervised data in training to improve TTS quality. It is usually achieved by transfer learning, i.e. pre-training TTS modules with only audio or text, and fine-tuning it using supervised data. For example, in Tacotron-based TTS \cite{wang2017tacotron, shen2018natural}, the auto-regressive decoder can be pre-trained with unlabeled audio \cite{chung2019semi, fujimoto2020semi}, then applied in supervised training to achieve better prediction quality. In \cite{Jia2021PnGBA}, the phoneme encoder is also pre-trained using only text in a BERT-like \cite{kenton2019bert} way. Besides, we can also tag unlabeled speech audio using limited supervised data. For example, speech chain \cite{tjandra2017listening}, i.e. back-translation \cite{caswell2019tagged}, can train Automatic Speech Recognition (ASR) and TTS iteratively using only a few paired data and much unpaired data. In \cite{inoue2020semi, kharitonov2023speak}, a weak ASR system is trained with a few minutes of supervised data, then decodes all unlabeled audio to generate transcripts to create a low-accuracy supervised dataset. TTS is directly pre-trained by this dataset, and then fine-tuned with the target supervised data. Besides, we can also create pseudo-labels via unsupervised acoustic unit discovery \cite{tu2020semi, kim2022transfer} to form a pseudo-supervised dataset for TTS pre-training. Eventually, employing more unlabeled data in training can enhance the TTS system with lower requirements for supervised data.

However, with the development of representation learning, data-derived speech representations show better performance than conventional, signal-processing-based acoustic features in TTS, which can be better predicted from the text while keeping sufficient speech information for high-quality audio reconstruction. It effectively reduces the requirement for supervised data, indicating a new direction for semi-supervised TTS.

\subsection{Self-Supervised Speech Representation Learning} 

S3RL \cite{mohamed2022self} aims to learn useful representations from unlabeled speech data to serve downstream supervised speech tasks. It is usually divided into two categories: contrastive and generative \cite{liu2021self}. Contrastive S3RL models, e.g. Wav2Vec \cite{schneider2019wav2vec, Baevski2020vq, baevski2020wav2vec}, HuBERT \cite{hsu2021hubert}, WavLM \cite{chen2022wavlm}, are usually encoder-based models trained with contrastive loss functions \cite{jaiswal2020survey}. This kind of model is more robust to noisy data in training, hence can be applied with massive low-quality speech data to learn a general speech representation \cite{feng2023superb} to enhance various downstream speech classification \cite{lee2022self, cao2021improving, chen2022large, zhu2022joint, zhao2022improving} and synthesis \cite{lim2021preliminary, wang2023comparative} tasks. Generative S3RL aims to reconstruct speech in training while applying restrictions to the latent space, which is more compatible with TTS intuitively due to their overlapped goal of speech generation. Hence, it is widely applied in TTS tasks, such as the Auto-regressive model, VAE \cite{burgess2018understanding}, and VQ-VAE \cite{Oord2017NeuralDR}. It cannot only provide an effective speech representation \cite{lu2021vaenar}, but provide a good pre-trained model for TTS training \cite{chung2019semi}. However, this kind of approach is more sensitive to noisy data due to the reconstruction objective \cite{liu2021self}, hence having a higher requirement for audio quality, lacking good generalization for low-quality audio.

\subsection{Vector-Quantized Representation Learning} 

As a generative self-supervised learning method, vector-quantized representation learning aims to compress the target via vector quantization into a compact, discrete representation, while keeping high-quality reconstruction, e.g. VQ-VAE \cite{Oord2017NeuralDR} and its enhanced version, VQ-GAN \cite{esser2021taming} and VQ-Diffusion \cite{gu2022vector}. It has been widely applied in speech generation, e.g. speech coding \cite{garbacea2019low, chen2021tenc, defossez2022high}, Voice Conversion (VC) \cite{wu2020one}, and TTS \cite{hayashi2020discretalk, guo2022msmc, wang2023neural}. To extract a better VQ speech representation with a balance between compactness and completeness, MSMC-TTS \cite{guo2022msmc} proposes a Multi-Stage Multi-Codebook Representation (MSMCR), comprising multiple sequences with different time resolutions and quantized by multiple codebooks. It can be better predicted from the text via multi-stage modeling, significantly improving TTS performance with lower supervised data requirements, further showing the great potential of VQ representations in semi-supervised TTS.

\begin{figure*}[htp]
    \centering
    \includegraphics[width=18cm]{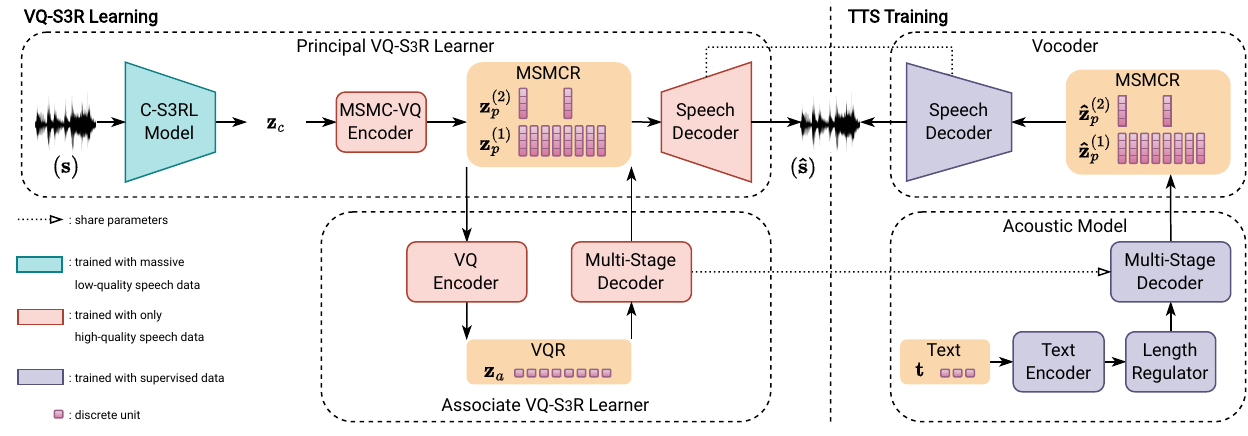}
    \caption{The framework of QS-TTS. In vector-quantized self-supervised speech representation learning (VQ-S3RL), the principal VQ-S3R learner first converts the speech signal $\mathbf{s}$ into $\mathbf{z}_c$ via a pre-trained contrastive S3RL model, and then feed it to the Multi-Stage Multi-Codebook (MSMC) VQ-GAN to obtain the MSMCR $\mathbf{z}_p$ and the reconstructed speech waveform $\mathbf{\hat{s}}$. The associate VQ-S3R learner compresses $\mathbf{z}_p$ into the VQ sequence $\mathbf{z}_a$ via a VQ-VAE model. In TTS training, the vocoder and acoustic model are trained based on the pre-trained speech decoder and multi-stage decoder to map the text to the MSMCR $\mathbf{\hat{z}}_p$, and then synthesize the waveform $\mathbf{\hat{s}}$.}
    \label{fig:qs_tts}
\end{figure*}

\section{QS-TTS}
\label{sec:model}

QS-TTS is a semi-supervised TTS framework based on vector-quantized self-supervised speech representation learning (VQ-S3RL). As shown in Fig. \ref{fig:qs_tts}, it trains two VQ-S3R learners on unlabeled speech data to provide more-profitable speech representations and effective pre-trained models to enhance supervised TTS training, thereby improving synthesis quality while reducing the supervised data requirement. In this section, we will illustrate each module in detail.

\subsection{The Principal VQ-S3R Learner}

We first propose a principal VQ-S3R learner combining contrastive and generative S3RL models to extract an effective speech representation, which is easier to predict from the text and well-reconstructed into high-quality audio. It first employs a contrastive S3RL model, HuBERT \cite{hsu2021hubert}, trained with massive speech audio to extract an effective general speech representation $\mathbf{z}_c$ from the speech signal $\mathbf{s}$. Then, it conducts the generative VQ-S3RL based on MSMC-VQ-GAN using only high-quality speech data to convert $\mathbf{z}_c$ into the generative VQ-S3R, MSMCR, while decoding it to high-quality audio $\mathbf{\hat{s}}$. In this section, we will introduce the model architecture and training method of MSMC-VQ-GAN. 

\subsubsection{Model architecture}

\begin{figure*}[htp]
    \centering
    \includegraphics[width=18cm]{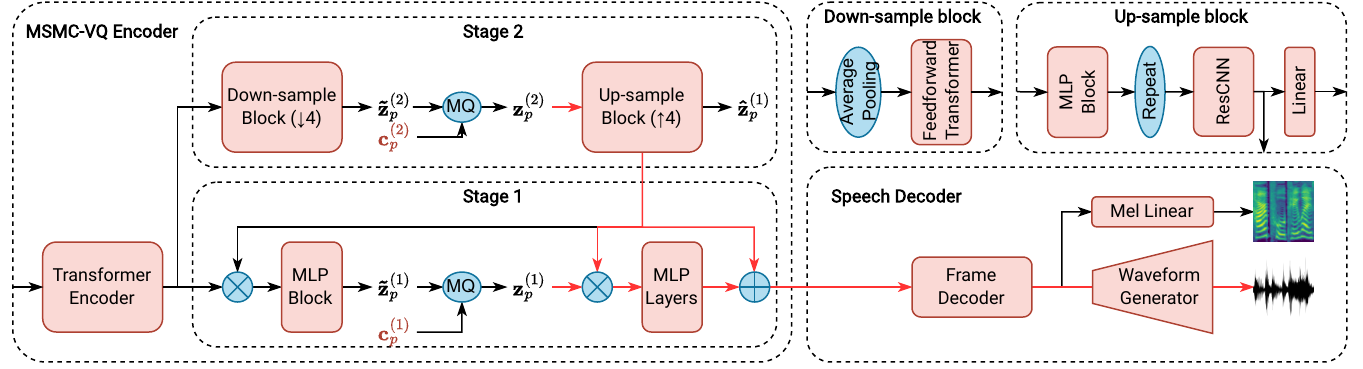}
    \caption{The model architecture of a two-stage four-codebook VQ-GAN generator. ``MQ'', ``$\bigoplus$'', ``$\bigotimes$'' denote multi-head vector quantization, addition, and concatenation operations. The red arrows emphasize the process of converting the MSMCR into the waveform at inference.}
    \label{fig:msmc_vqgan}
\end{figure*}

The model architecture of MSMC-VQ-GAN is composed of an MSMC-VQ encoder and a speech decoder. Fig. \ref{fig:msmc_vqgan} shows an example of a two-stage four-codebook VQ-GAN model. In the MSMC-VQ encoder, the input $\mathbf{z}_c$ is first processed by a Transformer encoder composed of a linear layer and a feedforward Transformer block, then quantized in two stages. In the higher stage, the input sequence is down-sampled 4 times along the time axis via the down-sample module, which has an average pooling layer and a feedforward Transformer block. Then, $\mathbf{\tilde{z}}_p^{(2)}$ is quantized by a four-head codebook $\mathbf{c}_p^{(2)}$ via Multi-Head Vector Quantization (MHVQ) \cite{guo2022msmc}, i.e. product quantization \cite{jegou2010product}, which chunks the codebook into multiple sub-codebooks to quantize the input vector chunked in the same way, respectively. The quantized output $\mathbf{z}_p^{(2)}$ is further processed by an up-sample block comprising two MLP layers with a LeakyReLU activation function in between, repetition operation for up-sampling, and four residual convolutional layers, to help the following quantization, and predict the lower-stage quantized sequence $\mathbf{\hat{z}}_p^{(1)}$. In stage 1, we obtain the quantized sequence $\mathbf{z}_p^{(1)}$ with the guidance of the high-stage information, and add it with the high-stage residual output as the encoder output, which is then fed to the speech decoder to generate the waveform. The speech decoder is composed of a frame decoder and a waveform generator. First, we employ the frame decoder with a feedforward Transformer block to process the whole input sequence, and then generate the speech waveform $\mathbf{\hat{s}}$ via the waveform generator based on the Hifi-GAN model. Meanwhile, like MSMC-VQ-VAE, we still predict the Mel spectrogram from the output of the frame decoder using a Mel linear layer.

\subsubsection{Loss function}

The training objective of MSMC-VQ-GAN is composed of multiple loss terms. First, due to the non-differentiable VQ operations, we back-propagate the gradient to the MSMC-VQ encoder via the following loss:
\begin{equation}
\label{eq:msvq}
    \mathcal{L}_{vq} = \frac{1}{S}\sum^{S}_{i=1} || \mathbf{\tilde{z}}_p^{(i)} - sg(\mathbf{z}_p^{(i)}) ||^2_2
\end{equation}
where $S$ denotes the number of stages. And we adopt the exponential moving average-based method \cite{razavi2019generating} to update codebooks in training. For effective multi-stage representation learning, we also enhance the relationship between adjacent stages with the following loss term:
\begin{equation}
    \mathcal{L}_{ms} = \frac{1}{S-1}\sum^{S-1}_{j=1} ||\mathbf{\hat{z}}_p^{(j)} - sg(\mathbf{z}_p^{(j)})||^2_2
\end{equation}
It can help the higher stage learn an effective representation stably \cite{guo2022msmc, Child2021VeryDV}. 

To reconstruct high-quality speech audio, we apply GAN training to the model with a UnivNet discriminator \cite{jang2021univnet}, composed of multiple sub-discriminators for multi-resolution spectrogram discriminating and multi-period waveform discriminating, to capture more discriminative information in frequency and time domains. The loss function for the discriminator is written as:
\begin{equation}
    \mathcal{L}_d = \frac{1}{K}\sum_{k=1}^K[(D_k(\mathbf{s})-1)^2 + D_k(\mathbf{\hat{s}})^2]
\end{equation}
where $K$ denotes the number of sub-discriminators. And the adversarial loss for the MSMC-VQ-GAN is written as:
\begin{equation}
    \mathcal{L}_{adv} = \frac{1}{K}\sum_{k=1}^K[(D_k(\mathbf{\hat{s}}) - 1)^2]
\end{equation}
To enhance GAN training quality, the Mel-spectrogram loss and feature matching loss, widely used in GAN-based neural vocoder training, are also employed in MSMC-VQ-GAN training \cite{hifigantts}. Mel-spectrogram loss is the L1 distance between two waveforms in the Mel-scale frequency domain, which can improve the perceptual quality of the generated audio. It is written as follows:
\begin{equation}
    \mathcal{L}_{mel} = ||\phi(\mathbf{s}) - \phi(\mathbf{\hat{s}})||_1
\end{equation}
where $\phi$ denotes the operating converting the waveform into the log-scale Mel spectrogram. Feature matching loss can further improve GAN training quality by reducing differences between the ground-truth waveform and the generated waveform in the hidden feature space of the discriminator as follows:
\begin{equation}
    \mathcal{L}_{fm} = \frac{1}{K}\sum_{k=1}^K \frac{1}{N_k}\sum_{i=1}^{N_k}||(D_k^{(i)}(\mathbf{s})) - D_k^{(i)}(\mathbf{\hat{s}})||_1
\end{equation}
where $N_k$ denotes the number of hidden layers of $k$-th sub-discriminator, and $D_k^{(i)}(*)$ denotes the output feature of $i$-th layer of $k$-th sub-discriminator.

Finally, to avoid the negative impact of the unstable performance of GAN to VQ-S3RL in the early-stage training, we still apply the frame-level reconstruction loss, which is an L2 distance between the predicted Mel spectrogram and the ground-truth one:
\begin{equation}
    \mathcal{L}_{frame} = || \mathbf{x} - \mathbf{\hat{x}} ||^2_2
\end{equation}
where $\mathbf{x}$ and $\mathbf{\hat{x}}$ denote the ground-truth and predicted Mel spectrograms. Finally, the loss function for MSMC-VQ-GAN is written as follows:
\begin{equation}
\begin{split}
    \mathcal{L}_{g} &= \mathcal{L}_{adv} + \lambda_{fm} * \mathcal{L}_{fm}  + \lambda_{mel} * \mathcal{L}_{mel} \\
    &+ \lambda_{vq} * \mathcal{L}_{vq} + \lambda_{ms} * \mathcal{L}_{ms} + \lambda_{frame} * \mathcal{L}_{frame} 
\end{split}
\end{equation}
where $\lambda_{fm}, \lambda_{mel}, \lambda_{vq}, \lambda_{ms}, \lambda_{frame}$ are weight coefficients. 

\subsection{The Associate VQ-S3R Learner}

To better predict the MSMCR from the text, we also propose an associate VQ-S3R learner to provide an effective pre-trained model for training the acoustic model. It imitates the process converting the text, a highly-compact discrete sequence, into the MSMCR, by employing a VQ-VAE-based model to abstract the MSMCR into a more compact VQ sequence and reconstruct it back.

\begin{figure}[htp]
    \includegraphics[width=8.5cm]{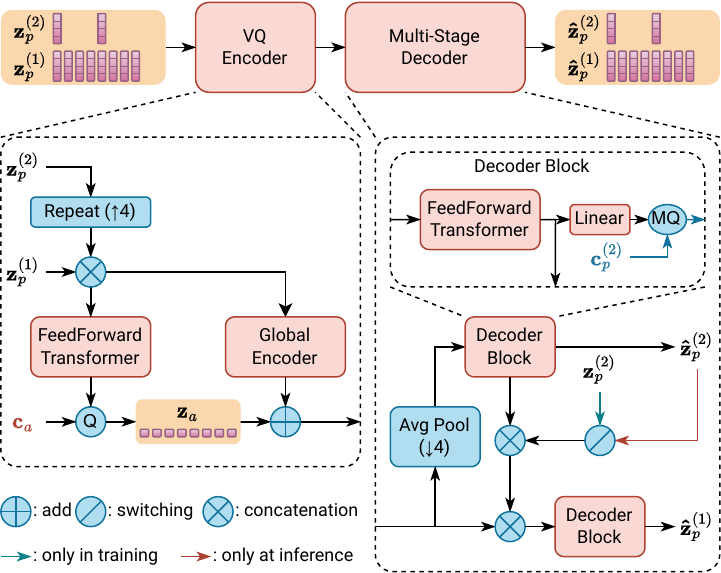}
    \caption{The model structure of VQ-VAE for the two-stage four-codebook representation, where the second-stage sequence has a down-sample rate of 4 along the time axis.}
    \label{fig:weak_vqs3rl}
\end{figure}

\subsubsection{Model architecture}

Fig. \ref{fig:weak_vqs3rl} shows the VQ-VAE model for the two-stage representation, where the second-stage sequence $\mathbf{z}_p^{(2)}$ has a down-sample rate of 4. It is composed of a VQ encoder and a multi-stage decoder. In the VQ encoder, all sequences are first up-sampled to the same length as the lowest-stage sequence $\mathbf{z}_p^{(1)}$, e.g. repeating all vectors in $\mathbf{z}_p^{(2)}$ for 4 times. Then, aligned sequences are concatenated together to be processed by a feedforward Transformer block, and quantized by one codebook to obtain a highly-compact VQ sequence $\mathbf{z}_a$. To force it to capture more phonetics-related information, we also employ a global encoder based on ECAPA-TDNN \cite{desplanques2020ecapa} to extract an utterance-level embedding representing some global attributes, e.g. speaker information, from the MSMCR. This embedding is then up-sampled to the frame level by repetition, and added with the quantized sequence $\mathbf{z}_a$ for decoding.

The multi-stage decoder aims to reconstruct the MSMCR from $\mathbf{z}_a$. It predicts sequences from high to low stages in a cascaded way. For this two-stage representation, the decoder first down-samples the encoder output for 4 times, then feeds it to the decoder block comprising a feedforward Transformer block followed by a linear layer. The output of the linear layer directly refers to the predicted sequence $\mathbf{\hat{z}}_p^{(2)}$ in training, but is quantized by the codebook $\mathbf{c}_p^{(2)}$ from the MSMC-VQ-GAN as the output. Similarly, in stage 1, $\mathbf{\hat{z}}_p^{(1)}$ is predicted by given the encoder output and the higher-stage outputs. Notably, in training, we replace $\mathbf{\hat{z}}_p^{(2)}$ with the ground-truth quantized sequence $\mathbf{z}_p^{(2)}$ as the input.

\subsubsection{Loss function}
 
The model is trained with the loss function $\mathcal{L}_{a}$ written as follows:
\begin{equation}
\label{eq:wvqs3l}
\begin{split}
    \mathcal{L}_{vq} &= || \mathbf{\tilde{z}}_a - sg(\mathbf{z}_a) ||^2_2 \\
    \mathcal{L}_{rec} &= \frac{1}{S}\sum^{S}_{i=1} ||\mathbf{\hat{z}}^{(i)}_p - \mathbf{z}^{(i)}_p||^2_2 \\
    \mathcal{L}_{a} &= \mathcal{L}_{vq} + \lambda_{rec} * \mathcal{L}_{rec}
\end{split}
\end{equation}
where $\mathcal{L}_{vq}$ is still the VQ loss for the VQ encoder training, $\mathcal{L}_{rec}$ is the reconstruction loss between the ground-truth and reconstructed MSMCR, and $\lambda_{rec}$ is a weight coefficient. In training, we still use the exponential moving average-based method to update the codebook $\mathbf{c}_a$.

\subsection{TTS Synthesis}

In TTS synthesis, we aim to convert the text to its corresponding MSMCR via the acoustic model, and then generate the waveform from the MSMCR via the vocoder. The acoustic model has the same architecture as MSMC-TTS \cite{guo2022msmc}, based on the FastSpeech \cite{FastSpeech}. It encodes the text sequence using a Transformer encoder, then up-samples it via repetition according to the predicted duration, finally generating the MSMCR $\mathbf{\hat{z}}_q$ via a multi-stage decoder. In training, it inherits parameters of the pre-trained multi-stage decoder in the associate VQ-S3R learner, and then is trained with the supervised data. The training loss function is written as follows:
\begin{equation}
\begin{split}
    \mathcal{L}_{dur} &= || \mathbf{\hat{d}} - \mathbf{d} ||^2_2 \\
    \mathcal{L}_{am} &= \mathcal{L}_{rec} + \lambda_{dur} * \mathcal{L}_{dur}
\end{split}
\end{equation}
where $\mathcal{L}_{dur}$ denotes the duration loss between the ground-truth duration $\mathbf{d}$ and the predicted duration $\mathbf{\hat{d}}$, and $\lambda_{dur}$ is a weight coefficient. 

In vocoder, to convert the predicted $\mathbf{\hat{z}}_p$ composed of multiple sequences to the waveform, we first feed it to the MSMC-VQ encoder in the pre-trained MSMC-VQ-GAN to obtain the encoder output sequence, as indicated by the red arrows shown in Fig. \ref{fig:msmc_vqgan}. Then, we synthesize the waveform via the speech decoder fine-tuned with the audio in the supervised dataset to adapt the target speaker better. In fine-tuning, we only update the parameters of the speech decoder using the same training configurations of MSMC-VQ-GAN.

\section{Experimental Protocol}
\label{sec:exp}

\subsection{Dataset}

In VQ-S3RL, we employ AIShell-3, a Mandarin multi-speaker high-quality speech dataset, as the training set. This dataset contains roughly 85 hours of emotion-neutral recordings spoken by 218 native Chinese Mandarin speakers, showing a rich coverage of phonetic and speaker information. In TTS training, we employ multiple supervised TTS datasets to evaluate TTS systems under various scenarios. The first dataset is \emph{CSMSC\footnote{CSMSC is available at \url{https://www.data-baker.com/data/index/source}.}}, a single-speaker Mandarin TTS corpus with 10-hour high-quality supervised data, which is widely applied in training the standard single-speaker Mandarin TTS system. We also conduct low-resource scenarios by extracting subsets from CSMSC. Besides, a test set with 200 utterances is also extracted from this dataset but has no overlap with all training sets. Then, we construct a more challenging low-resource scenario using only 10 minutes of child speech data spoken by a five-year-old girl in Mandarin\footnote{The child TTS dataset is available at \url{https://magichub.com/datasets/mandarin-chinese-speech-corpus-for-tts-children-speech}.}. It has a test set with 24 utterances out of the training set. Finally, we use an internal Cantonese dataset with 15 minutes of supervised data to evaluate the performance of TTS systems in low-resource languages. The Cantonese test set comprises 134 utterances out of the training set.

\subsection{Feature}

First, in audio processing, all single-channel audio used in our work is down-sampled to the sample rate of 16kHz. Then, we extract Mel spectrograms for all datasets in the following way: first, pre-emphasize the audio with the coefficient of 0.97, and then convert it to the 1025-dim magnitude spectrograms by STFT with a window length of 50ms, a frameshift of 12.5ms, and an FFT size of 2048, finally compress the spectrogram into the 80-dim log-scale Mel spectrogram. We employ a HuBERT\footnote{The pre-trained HuBERT model is available at \url{https://huggingface.co/TencentGameMate/chinese-hubert-large}.} \cite{hsu2021hubert}, the contrastive S3RL model, pre-trained on WenetSpeech \cite{zhang2022wenetspeech}, a Mandarin dataset with around 10,000 hours of speech audio, to extract the general speech representation. The HuBERT feature is a sequence of 1024-dim vectors with a frameshift of 20ms. To align it to the Mel spectrogram, we up-sample it via nearest neighbor interpolation to a frameshift of 12.5ms.

In text processing, we convert the text to phonemes as the input of the acoustic model. For CSMSC and child datasets, we directly employ the phonemes and their corresponding duration labeled in the dataset for training. For the Cantonese dataset, we use an open-source G2P tool \cite{lee-etal-2022-pycantonese} to obtain phonemes, and train a Montreal Forced Aligner\footnote{The tool is available at \url{https://github.com/MontrealCorpusTools/Montreal-Forced-Aligner}.} (MFA) model to obtain the phoneme-level duration for training.

\subsection{Model Configuration}

In VQ-S3RL, we apply the MSMC-VQ-GAN with 2 stages, where the second stage has a down-sample rate of 4 along the time axis. And in each stage, a 4-head codebook is used for vector quantization, where each head is composed of 64 codewords with a dimension of 64. In the MSMC-VQ encoder, we apply the 4-layer 256-dim feedforward Transformer block with 2-head self-attention to the Transformer encoder and the down-sample block. The MLP block has two 256-dim linear layers with a Tanh activation function. The residual CNN block comprises 4 1-D residual convolutional layers with a kernel size of 5. In the speech decoder, the frame decoder is also implemented with the 4-layer 256-dim feedforward Transformer block. The waveform generator is a Hifi-GAN-V1 \cite{hifigantts} generator, which upsamples the input sequence 200 times to the 16kHz waveform via 4 CNN-based upsampling blocks with the upsample rates of $[5, 5, 4, 2]$ and the kernel sizes of $[11, 11, 8, 4]$. 

In GAN training based on the UnivNet discriminator (UnivNet-c32) \cite{jang2021univnet}, we extract three magnitude spectrograms from the waveform using three STFT parameter sets, FFT size $[256, 512, 1024]$, frameshift $[40, 80, 160]$, and frame length $[120, 320, 640]$, for multi-resolution spectrogram discriminating. And we also reshape the 1-D waveform into five 2-D sequences with the period of $[2, 3, 5, 7, 11]$ for multi-period waveform discriminating\footnote{The implementations of Hifi-GAN generator and UnivNet discriminator are available at \url{https://github.com/jik876/hifi-gan} and \url{https://github.com/mindslab-ai/univnet}.}. In this work, MSMC-VQ-GAN is trained on AIShell-3 for 400k iterations using the AdamW optimizer ($\beta_1=0.8, \beta_2=0.99$) with a batch size of 16 utterances. Similar to random window discriminating \cite{donahue2021endtoend}, we also randomly select a segment with a length of 0.75 seconds from each utterance for adversarial training of the waveform generator to improve training efficiency. The learning rate of $2 \times 10^{-4}$ exponentially decays with the rate $2^{-\frac{1}{200,000}}$ after 200k warm-up iterations. The weight coefficients $\lambda_{fm}, \lambda_{mel}, \lambda_{vq}, \lambda_{ms}, \lambda_{frame}$ are set to 2, 45, 10, 1, 450, respectively. To stabilize the training process, we also apply warm-up training here, i.e. no GAN training in the first 50k iterations.

The associate VQ-S3R learner compresses the MSMCR with one codebook with 64 256-dim codewords. The VQ encoder is a Transformer encoder with the same configuration as MSMC-VQ-GAN. And the global encoder is an ECAPA-TDNN \cite{desplanques2020ecapa} model with 128 channels for hidden layers. This model is trained on AIShell-3 for 200k iterations using Adam \cite{KingmaB14} optimizer ($\beta_1=0.9, \beta_2=0.98$) with the batch size of 64 utterances, and the learning rate of $2 \times 10^{-4}$ exponentially decayed with the rate $2^{-\frac{1}{20,000}}$ after 20k warm-up iterations.

Finally, in TTS training, we use the supervised TTS dataset to fine-tune the speech decoder of MSMC-VQ-GAN with the same training configuration, and train the acoustic model with $\lambda_{dur} = 0.1$ based on the pre-trained multi-stage decoder in the associate VQ-S3R learner. For a standard supervised dataset with sufficient audio, i.e. 10-hour CSMSC, we fine-tune the vocoder for 400k iterations and the acoustic model for 200k iterations. Otherwise, for all low-resource datasets, we only fine-tune the vocoder for 100k iterations, and the acoustic model for 50k iterations to avoid over-fitting.

\subsection{Baselines}

In our experiments, we implement multiple fully-supervised and semi-supervised TTS approaches following the training configuration of QS-TTS. 

\subsubsection{FastSpeech}
It is a mainstream non-autoregressive neural TTS system based on the Mel spectrogram. It comprises an acoustic model based on Transformer blocks and a Hifi-GAN vocoder, having the same model hyperparameters as QS-TTS. We upgrade it to two semi-supervised versions: FastSpeech-S and FastSpeech-SS. In FastSpeech-S, the Mel-spectrogram-based Hifi-GAN vocoder is pre-trained on AI-Shell3, and then fine-tuned with the supervised dataset. In FastSpeech-SS, the Mel spectrogram is replaced with HuBERT features, and the HuBERT-based Hifi-GAN vocoder is also pre-trained on AI-Shell3.

\subsubsection{VITS \cite{kim2021conditional}}
It is the SOTA end-to-end TTS system based on VAE representations. It applies a VAE-GAN model based on Hifi-GAN to learn a generative speech representation, which can be predicted from the text by a Glow-based module \cite{glowtts}. We implement it based on the official model configuration\footnote{The official implementation of VITS is available at \url{https://github.com/jaywalnut310/vits}}, and apply it with the waveform generator and the discriminator with the same configuration as that of QS-TTS for a fair comparison. We also implement its semi-supervised version, VITS-SS \cite{kim2022transfer}, in our experiments. It first applies k-means clustering with 512 centroids on HuBERT features of AIShell-3 to create pseudo-labels for all unlabeled audio, then pre-trains VITS on this created paired dataset, and finally fine-tunes VITS using the supervised dataset.

\subsubsection{MSMC-TTS}
It can be seen as the fully-supervised version of QS-TTS, which learns the MSMCR from the Mel spectrogram via the MSMC-VQ-GAN model, and is trained with only the supervised dataset. Besides, We also implement another semi-supervised version of MSMC-TTS, MSMC-TTS-SS, which pre-trains the Mel-spectrogram-based MSMC-VQ-GAN on AIShell-3, and only fine-tunes the speech decoder in TTS training.

\subsubsection{Back Translation}
We also implement this semi-supervised training approach to pre-train the acoustic model. It first employs the supervised TTS dataset to train an ASR system, and then transcribe unlabeled speech audio of AIShell-3 to create a low-precision but large-scale paired dataset. The acoustic model is pre-trained on this dataset, and finally fine-tuned with the supervised TTS dataset. This approach highly relies on the quality of ASR to ensure precise transcription, so we train a CTC-ASR based on the pre-trained Mandarin HuBERT model to convert the audio into phonemes and their corresponding durations extracted from CTC alignments. Besides, we also implement the k-means-based approach proposed in VITS-SS to avoid training ASR models. It directly quantize HuBERT features of AIShell-3 into 512 codewords by k-means to obtain pseudo-labels and corresponding durations, which are then applied to pre-train the acoustic model.

\subsection{Evaluation}

In this work, we conduct objective and subjective tests to evaluate the proposed TTS approach comprehensively.

\subsubsection{Objective Metrics}

Except for Mel Cepstral Distortion (MCD) \cite{kubichek1993mel}, computing the perceptual difference between two fully-aligned audio files, we also propose evaluating the synthesis quality using Frechèt distances \cite{heusel2017gans} in various embedding spaces. Frechèt distance can measure the distance between two sets of samples by calculating the difference between them in distributions as follows:
\begin{equation}
    \mathbb{F}(\mathcal{N}_s, \mathcal{N}_t) = || \mu_s - \mu_t ||^2 + tr(\Sigma_s + \Sigma_t - 2 \sqrt{\Sigma_s \Sigma_t})
\end{equation}
where $\mathcal{N}_s$ and $\mathcal{N}_t$ denote the normal distribution of the synthesized audio set, and the target ground-truth audio set in the embedding space. And we can use different audio classification models to extract embeddings and calculate their mean vectors ($\mu_s, \mu_t$) and covariance matrices ($\Sigma_s, \Sigma_t$). This work employs three embedding spaces: acoustics, speaker, and phonetics. The Frechèt distance in acoustic space, i.e. Frechèt Audio Distance (FAD) \cite{kilgour2018fr}, has been well applied in evaluating the synthesis quality of neural vocoder, which extracts the embedding for each 4-second audio from an audio classification model. Similarly, we extract utterance-level speaker embeddings using an ECAPA-TDNN-based speaker verification model, and extract utterance-level phonetic embeddings using a Transformer-based ASR model by averaging its encoder output sequence into one vector. These three distances are denoted as FD-AC, FD-SV, and FD-ASR in our following work\footnote{The pre-trained audio classification, speaker verification, and ASR models are available at \url{https://github.com/harritaylor/torchvggish}, \url{https://huggingface.co/speechbrain/spkrec-ecapa-voxceleb}, and \url{https://github.com/openai/whisper/tree/main} (multi-lingual base version).}. Notably, the Frechèt distance in speaker space is multiplied by 10 to align with other distances.

We also evaluate intelligibility, the most crucial factor in evaluating a TTS system, by calculating the Character Error Rate (CER) and Phoneme Error Rate (PER) by transcribing the synthesized audio using ASR tools\footnote{The ASR tools for Mandarin and Cantonese are available at \url{https://github.com/wenet-e2e/wenet/tree/main/runtime/binding/python} and \url{https://huggingface.co/Scrya/whisper-large-v2-cantonese}.}. And we use G2P tools\footnote{The G2P tools for Mandarin and Cantonese are available at \url{https://github.com/mozillazg/python-pinyin} and \url{https://github.com/mozillazg/python-pinyin}.} to convert the transcribed and ground-truth text into phonemes to calculate PER, which focuses more on the pronunciation accuracy of phonemes.

\subsubsection{Subjective Metrics}

We conduct MOS (mean opinion score) tests to subjectively evaluate TTS systems on synthesis quality. In each MOS test, 10 native speakers are hired to rate each audio sample in 20 test cases, where each test case contains multiple audio samples synthesized by different TTS approaches but from the same text. The rating ranges from 1 to 5 with an increment of 0.5, where the higher score indicates better quality. Finally, we statistic the scores of each method to obtain the MOS with a 95\% confidence interval.

\begin{table}[htp]
\centering
\caption{MOS test: Standard single-speaker Mandarin TTS}
\label{tab:mos1}
\begin{tabular}{c|ccc|cc|c}
\hline
\multirow{2}{*}{Systems} & \multicolumn{3}{c|}{FDs}                      & \multicolumn{2}{c|}{ERs (\%)}      & \multirow{2}{*}{\begin{tabular}[c]{@{}c@{}}MOS \\ ($\pm$ 95\%CI)\end{tabular}} \\ \cline{2-6}
                & AC   & SV   & ASR  & CER  & PER  &           \\ \hline
Recording       & -    & -    & -    & 4.70 & 0.67 & 4.33±0.10 \\ \hline
FastSpeech      & 0.79 & 1.19 & 1.52 & 6.28 & \textbf{0.70} & 3.55±0.09 \\
VITS            & 0.22 & 0.96 & 1.85 & 7.96 & 1.15 & 3.74±0.09 \\
MSMC-TTS        & 0.34 & 0.84 & 1.51 & 6.11 & 0.80 & 3.99±0.09 \\
\textbf{QS-TTS} & \textbf{0.19} & \textbf{0.81} & \textbf{1.47} & \textbf{6.05} & 0.80 & \textbf{4.00±0.09}   \\ \hline
\end{tabular}
\end{table}

\begin{table}[htp]
\centering
\caption{MOS test: Low-resource single-speaker Mandarin TTS}
\label{tab:mos2}
\begin{tabular}{c|ccc|cc|c}
\hline
\multirow{2}{*}{Systems} & \multicolumn{3}{c|}{FDs}                      & \multicolumn{2}{c|}{ERs (\%)}      & \multirow{2}{*}{\begin{tabular}[c]{@{}c@{}}MOS \\ ($\pm$ 95\%CI)\end{tabular}} \\ \cline{2-6}
           & AC   & SV   & ASR  & CER   & PER  &           \\ \hline
Recording  & -    & -    & -    & 4.70  & 0.67 & 4.51±0.10 \\ \hline
FastSpeech & 4.17 & 1.79 & 4.10 & 12.28 & 3.23 & 2.58±0.11 \\
VITS       & 0.92 & 1.52 & 3.19 & 19.56 & 6.31 & 2.86±0.11 \\
MSMC-TTS   & 0.54 & 1.40 & 4.44 & 16.80 & 4.99 & 3.15±0.11 \\
\textbf{QS-TTS}          & \textbf{0.25} & \textbf{1.00} & \textbf{2.20} & \textbf{8.98} & \textbf{1.46} & \textbf{3.75±0.10}   \\ \hline
\end{tabular}
\end{table}

\begin{figure*}[htp]
    \includegraphics[width=18cm]{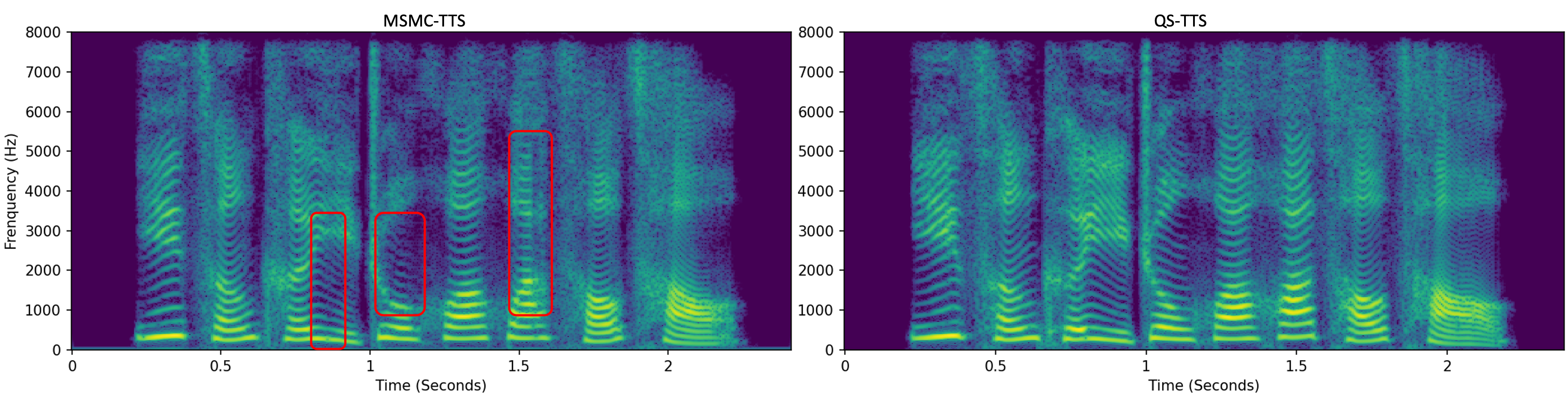}
    \caption{The magnitude spectrograms of audio samples synthesized from the same input text by MSMC-TTS (left) and QS-TTS (right) trained with 15-minute CSMSC. Red boxes highlight areas with significant differences.}
    \label{fig:tts_sample}
\end{figure*}

\section{Results}
\label{sec:result}

\subsection{TTS comparison: Semi-Supervised v.s. Supervised}

First, we compare the proposed semi-supervised TTS system with supervised TTS systems: FastSpeech, VITS, and MSMC-TTS. Table \ref{tab:mos1} shows the MOS test result of standard single-speaker TTS systems on the 10-hour CSMSC. FastSpeech, the modular TTS system based on the Mel Spectrogram, shows the worst synthesis quality among all TTS approaches, obtaining the lowest MOS of 3.55 and the highest FD-AC and FD-SV of 0.79 and 1.19. But it has a satisfying performance in pronunciation accuracy, showing the lowest PER of 0.70\%. VITS, the end-to-end approach based on VAE, significantly improves audio quality by a much lower FD-AC and FD-SV, and shows a higher MOS of 3.74. But the intelligibility also seriously degrades with the increased CER and PER. Instead, MSMC-TTS performs well in both intelligibility and synthesis quality, obtaining much-decreased metrics over FastSpeech, and a higher MOS of 3.99, which indicates the effectiveness of this approach based on VQ speech representations. QS-TTS inherits this approach, and is enhanced via the proposed VQ-S3RL using more unlabeled speech data. It further improves the overall performance, although sufficient supervised data is given in training. 

Then, we conduct this comparison in a more challenging TTS task, a low-resource scenario with only 15 minutes of supervised data. We randomly select 257 utterances from the full training set of CSMSC to form a 15-minute training set. As shown in Table \ref{tab:mos2}, all supervised approaches degrades seriously in audio quality and intelligibility under this scenario, achieving very low subjective scores. VITS and MSMC-TTS still achieve higher audio quality than FastSpeech, showing lower FD-AC and FD-SV. However, due to insufficient data to learn effective speech representations, their performance on intelligibility has more severe degradation, where the end-to-end approach, VITS, shows the highest CER and PER. QS-TTS addresses this problem by conducting VQ-S3RL on more unlabeled speech data, significantly improving audio quality and intelligibility. It shows significantly improved audio quality by an FD-AC of 0.25 and an FD-SV of 1.00, even lower than FastSpeech trained with 10 hours of supervised data. The intelligibility is also enhanced with the CER of 8.98 and PER of 1.46. Finally, its MOS of 3.75 further validates that QS-TTS surpasses all supervised methods greatly in this low-resource scenario.

As shown in Fig. \ref{fig:tts_sample}, we also visualize the magnitude spectrograms of samples synthesized by the MSMC-TTS and QS-TTS to investigate their differences further. The high-quality audio usually presents clear and smooth harmonics in the middle- and low-frequency parts to produce accurate pronunciation perceptually. However, MSMC-TTS trained with only 15 minutes of supervised data cannot synthesize the expected harmonics stably. It often presents a jittered pitch, and fuzzy middle-frequency harmonics, as shown in the red boxes, which leads to a degradation of audio quality and intelligibility. In QS-TTS, these issues are alleviated significantly. The synthesized audio shows clearer smoother harmonics in low- and middle-frequency parts, and has a higher variance in the high-frequency part, finally leading to a higher MOS in the subjective test.

\begin{table}[htp]
\centering
\caption{MOS Test: Different Semi-Supervised TTS approaches on the 10-minute child TTS dataset}
\label{tab:mos3}
\begin{tabular}{c|ccc|cc|c}
\hline
\multirow{2}{*}{Systems} & \multicolumn{3}{c|}{FDs}                      & \multicolumn{2}{c|}{ERs (\%)}      & \multirow{2}{*}{\begin{tabular}[c]{@{}c@{}}MOS \\ ($\pm$ 95\%CI)\end{tabular}} \\ \cline{2-6}
           & AC   & SV   & ASR  & CER  & PER  &           \\ \hline
Recording & - & - & - & 5.06 & 2.03 & 4.17±0.10 \\ \hline
MSMC-TTS & 1.76 & 3.37 & 7.49 & 27.53 & 13.02 & 2.93±0.13 \\
FastSpeech-SS & 1.72 & 3.31 & 7.73 & 25.28 & 15.48 & 2.80±0.12 \\
VITS-SS & \textbf{1.07} & 2.90 & 6.84 & 20.79 & 10.85 & 3.20±0.13 \\
\textbf{QS-TTS} & 1.48 & \textbf{2.81} & \textbf{6.32} & \textbf{14.61} & \textbf{5.35} & \textbf{3.23±0.14} \\ \hline
\end{tabular}
\end{table}

\begin{table}[htp]
\centering
\caption{MOS Test: Different Semi-Supervised TTS approaches on the 15-minute Cantonese TTS dataset}
\label{tab:mos4}
\begin{tabular}{c|ccc|cc|c}
\hline
\multirow{2}{*}{Systems} & \multicolumn{3}{c|}{FDs}                      & \multicolumn{2}{c|}{ERs (\%)}      & \multirow{2}{*}{\begin{tabular}[c]{@{}c@{}}MOS \\ ($\pm$ 95\%CI)\end{tabular}} \\ \cline{2-6}
           & AC   & SV   & ASR  & CER   & PER  &           \\ \hline
Recording & - & - & - & 8.53 & - & 3.86±0.11 \\ \hline
MSMC-TTS & 1.57 & 2.16 & 5.89 & 25.47 & 11.61 & 3.37±0.11 \\
FastSpeech-SS & \textbf{0.49} & 1.63 & 4.03 & 20.00 & 6.76 & 3.58±0.10 \\
VITS-SS & 0.57 & 1.67 & 5.20 & 42.35 & 19.92 & 3.34±0.11 \\
\textbf{QS-TTS} & 0.51 & \textbf{1.53} & \textbf{3.86} & \textbf{18.76} & \textbf{6.26} & \textbf{3.66±0.10} \\ \hline
\end{tabular}
\end{table}

\subsection{TTS Comparison: Semi-Supervised Approaches}

In the experiment, we compare QS-TTS with semi-supervised TTS systems, FastSpeech-SS and VITS-SS, to further validate the effectiveness of QS-TTS. First, we compare different semi-supervised TTS systems under an intra-lingual cross-style low-resource scenario using the 10-minute child speech dataset. It shares the same languages as AIShell-3, but has a unique timbre of a five-year-old girl unseen in AIShell-3. As shown in Table \ref{tab:mos3}, in this more challenging task with fewer supervised data, all approaches show higher Frechèt distances and error rates. First, the SOTA-supervised approach, MSMC-TTS, shows the worst intelligibility with the high CER and PER of 27.53\% and 13.02\%, and also performs poorly on audio quality with the highest FD-AC of 1.76. The semi-supervised version of FastSpeech, FastSpeech-SS, is significantly enhanced over its fully-supervised version, achieving comparable performance to MSMC-TTS. And VITS-SS enhanced by semi-supervised training outperforms MSMC-TTS significantly, achieving much lower Frechèt distances and error rates. Finally, although QS-TTS has a higher FD-AC than VITS-SS, it obtains the best performance on all other metrics, especially with a CER of 14.61\% of a PER of 5.35\% which are twice as low as those of MSMC-TTS. It indicates the universal effectiveness of QS-TTS in extreme scenarios.

Then, we further validate its effectiveness in a cross-lingual scenario, which builds a TTS system for a low-resource language, Cantonese, using a 15-minute supervised dataset, where the language is also unseen in the unlabeled speech dataset. As shown in Table \ref{tab:mos4}, this dataset has a relatively lower audio fidelity and lower expressiveness on prosody, hence achieving the MOS of 3.86 only. FastSpeech-SS shows a significant improvement over MSMC-TTS in this task, obtaining a higher MOS of 3.58, and lower metrics in audio quality and intelligibility. But VITS-SS fails to keep a stable performance, showing seriously degraded intelligibility with the much higher CER and PER of 42.35\% and 19.92\%. The pseudo-labels extracted from the Mandarin dataset using k-means cannot adapt to Cantonese well, providing a seriously biased pre-trained model. It leads to the lowest MOS of 3.34, showing the limitation of this semi-supervised approach in cross-lingual scenarios. However, QS-TTS still performs best with the highest MOS of 3.66. It still has the lowest CER and PER, and keeps high audio quality with comparable or lower Frechèt distances than other approaches.

In conclusion, through these challenging low-resource experiments, QS-TTS is validated as a more effective and stable approach in improving both audio quality and intelligibility of TTS over other supervised and semi-supervised approaches.

\begin{table*}[htp]
\centering
\caption{The performance of semi-supervised approaches based on different speech representations in analysis-synthesis and TTS synthesis using the 15-minute Mandarin TTS dataset.}
\label{tab:sr_tts}
\begin{tabular}{c|cc|cccccc|ccccc}
\hline
\multirow{3}{*}{TTS System} & \multicolumn{2}{c|}{Speech Representations} & \multicolumn{6}{c|}{Analysis-Synthesis} & \multicolumn{5}{c}{TTS Synthesis} \\ \cline{2-14} 
 & \multirow{2}{*}{HuBERT} & \multirow{2}{*}{MSMC-VQ} & \multicolumn{1}{c|}{\multirow{2}{*}{MCD (dB)}} & \multicolumn{3}{c|}{FDs} & \multicolumn{2}{c|}{ERs (\%)} & \multicolumn{3}{c|}{FDs} & \multicolumn{2}{c}{ERs (\%)} \\ \cline{5-14} 
 &  &  & \multicolumn{1}{c|}{} & AC & SR & \multicolumn{1}{c|}{ASR} & CER & PER & AC & SR & \multicolumn{1}{c|}{ASR} & CER & PER \\ \hline
FastSpeech-S & $\times$ & $\times$ & \multicolumn{1}{c|}{2.23} & 0.12 & 0.17 & \multicolumn{1}{c|}{0.15} & 4.61 & 0.66 & 2.29 & 1.51 & \multicolumn{1}{c|}{3.52} & 10.54 & 2.47 \\
MSMC-TTS-SS & $\times$ & $\checkmark$ & \multicolumn{1}{c|}{3.43} & 0.23 & 0.66 & \multicolumn{1}{c|}{0.52} & 5.61 & 0.99 & 0.74 & 1.95 & \multicolumn{1}{c|}{3.07} & 11.98 & 2.83 \\
FastSpeech-SS & $\checkmark$ & $\times$ & \multicolumn{1}{c|}{3.09} & 0.12 & 0.45 & \multicolumn{1}{c|}{0.30} & 4.79 & 0.58 & 0.43 & 1.10 & \multicolumn{1}{c|}{2.73} & 10.87 & 2.19 \\
QS-TTS-P & $\checkmark$ & $\checkmark$ & \multicolumn{1}{c|}{3.32} & 0.15 & 0.59 & \multicolumn{1}{c|}{0.41} & 5.11 & 0.79 & \textbf{0.34} & \textbf{1.03} & \multicolumn{1}{c|}{\textbf{2.34}} & \textbf{9.31} & \textbf{1.72} \\ \hline
\end{tabular}
\end{table*}

\begin{table*}[htp]
\centering
\caption{The TTS performance of acoustic models based on different pre-training methods in Mandarin and Cantonese TTS with 15-minute supervised training data. ``-'' denotes that no pre-training is applied to the acoustic model.}
\label{tab:tl}
\begin{tabular}{c|ccccc|ccccc}
\hline
\multirow{3}{*}{Methods} & \multicolumn{5}{c|}{Mandarin} & \multicolumn{5}{c}{Cantonese} \\ \cline{2-11} 
 & \multicolumn{3}{c|}{FDs} & \multicolumn{2}{c|}{ERs (\%)} & \multicolumn{3}{c|}{FDs} & \multicolumn{2}{c}{ERs (\%)} \\ \cline{2-11} 
 & AC & SR & \multicolumn{1}{c|}{ASR} & CER & PER & AC & SR & \multicolumn{1}{c|}{ASR} & CER & PER \\ \hline
- & 0.34 & 1.03 & \multicolumn{1}{c|}{2.34} & 9.31 & 1.72 & 0.58 & 1.62 & \multicolumn{1}{c|}{4.07} & 20.39 & 7.23 \\
Back Translation (ASR) & \textbf{0.25} & \textbf{0.99} & \multicolumn{1}{c|}{\textbf{2.11}} & \textbf{7.58} & \textbf{1.09} & 0.62 & 1.58 & \multicolumn{1}{c|}{4.00} & 21.63 & 7.89 \\
Back Translation (k-means) & 0.25 & 1.03 & \multicolumn{1}{c|}{2.37} & 10.77 & 2.30 & 0.52 & 1.55 & \multicolumn{1}{c|}{4.10} & 23.97 & 9.11 \\ 
Proposed & 0.25 & 1.00 & \multicolumn{1}{c|}{2.20} & 8.98 & 1.46 & \textbf{0.51} & \textbf{1.53} & \multicolumn{1}{c|}{\textbf{3.86}} & \textbf{18.76} & \textbf{6.26} \\ \hline
\end{tabular}
\end{table*}

\subsection{The Principal VQ-S3R Learner}

We conduct experiments for two VQ-S3R learners respectively to investigate the impact of the proposed VQ-S3RL on TTS. The principal learner combines HuBERT-based contrastive S3RL and MSMC-VQ-GAN-based generative S3RL to benefit TTS maximally. To validate the effectiveness of these two components, we implement QS-TTS-P, i.e. QS-TTS with only the principal learner, and compare it with the following semi-supervised TTS systems: FastSpeech-S, FastSpeech-SS, and MSMC-TTS-SS. All of these systems use AIShell-3 for pre-training and the 15-minute supervised data of CSMSC for TTS training.

As shown in Table \ref{tab:sr_tts}, first, The Mel-spectrogram-based FastSpeech-S, which does not use HuBERT and MSMC-VQ, performs best in analysis-synthesis, since the Mel spectrogram contains sufficient acoustic information for complete speech reconstruction. However, this feature with abundant information is hard to predict by the acoustic model without sufficient supervised data for training, leading to the worst TTS performance on audio quality. In MSMC-TTS-SS, MSMC-VQ compressed the Mel spectrogram into a more compact representation with less information. This lossy compression degrades the analysis-synthesis quality, but makes the feature easier to predict from the model trained with less supervised training data. It leads to a smaller gap between the ground-truth and predicted features, making TTS synthesis closer to analysis-synthesis in audio quality. However, the information loss also degrades intelligibility, causing a trade-off between audio quality and intelligibility in TTS synthesis. Hence, it is not advisable to overly compress features to enhance TTS unless we can keep sufficient phonetic information in compression. In FastSpeech-SS, the HuBERT, a contrastive speech representation learned from massive speech audio, discards more acoustic information, showing a lower audio quality in analysis-synthesis, but keeps richer phonetic information, which has a lower PER of 0.58\% than that of the Mel spectrogram, validates its completeness in representing speech. This feature also significantly improves TTS in both audio quality and intelligibility. Finally, QS-TTS-P applies MSMC-VQ with HuBERT to learn the compact representation with rich phonetic information, and achieves the best performance in TTS, strongly verifying the effectiveness of the principal VQ-S3R learner in TTS.

\subsection{The Associate VQ-S3R Learner}

The associate VQ-S3R learner aims to provide a practical pre-trained model to enhance the acoustic model to predict the MSMCR better. To validate its effectiveness on TTS, as shown in Table. \ref{tab:tl}, we compare it with another pre-training approach, back translation, under low-resource scenarios using 15 minutes of supervised data in Mandarin or Cantonese. 
 
First, the baseline approach, ASR-based back-translation, does not achieve consistent performance in these two scenarios. In Mandarin TTS, the ASR enhanced by Mandarin-HuBERT and trained with the Mandarin TTS dataset can transcribe the Mandarin unlabeled speech dataset well, providing a good pre-training paired dataset with sufficient transcription precision to support model pre-training, achieving the best TTS quality among all methods. However, in Cantonese TTS, the ASR trained with the Cantonese TTS dataset cannot transcribe Mandarin speech audio well into Cantonese phonemes, leading to the low-quality paired dataset. The pre-trained acoustic model on this dataset contaminates the following fine-tuning with Cantonese supervised data instead, showing higher CER and PER. It indicates the limitation of the ASR-based back-translation under the cross-lingual application. Although the k-means-based back-translation avoids ASR training, and shows consistent performance in both scenarios, it only improves the audio quality while degrading intelligibility. Different from these back-translation-based approaches, the associate VQ-S3R learner can enhance TTS in both audio quality and intelligibility under intra-lingual and cross-lingual scenarios, hence validated as an effective and general pre-training approach.

\begin{figure}[htp]
\centering
\includegraphics[width=8.8cm]{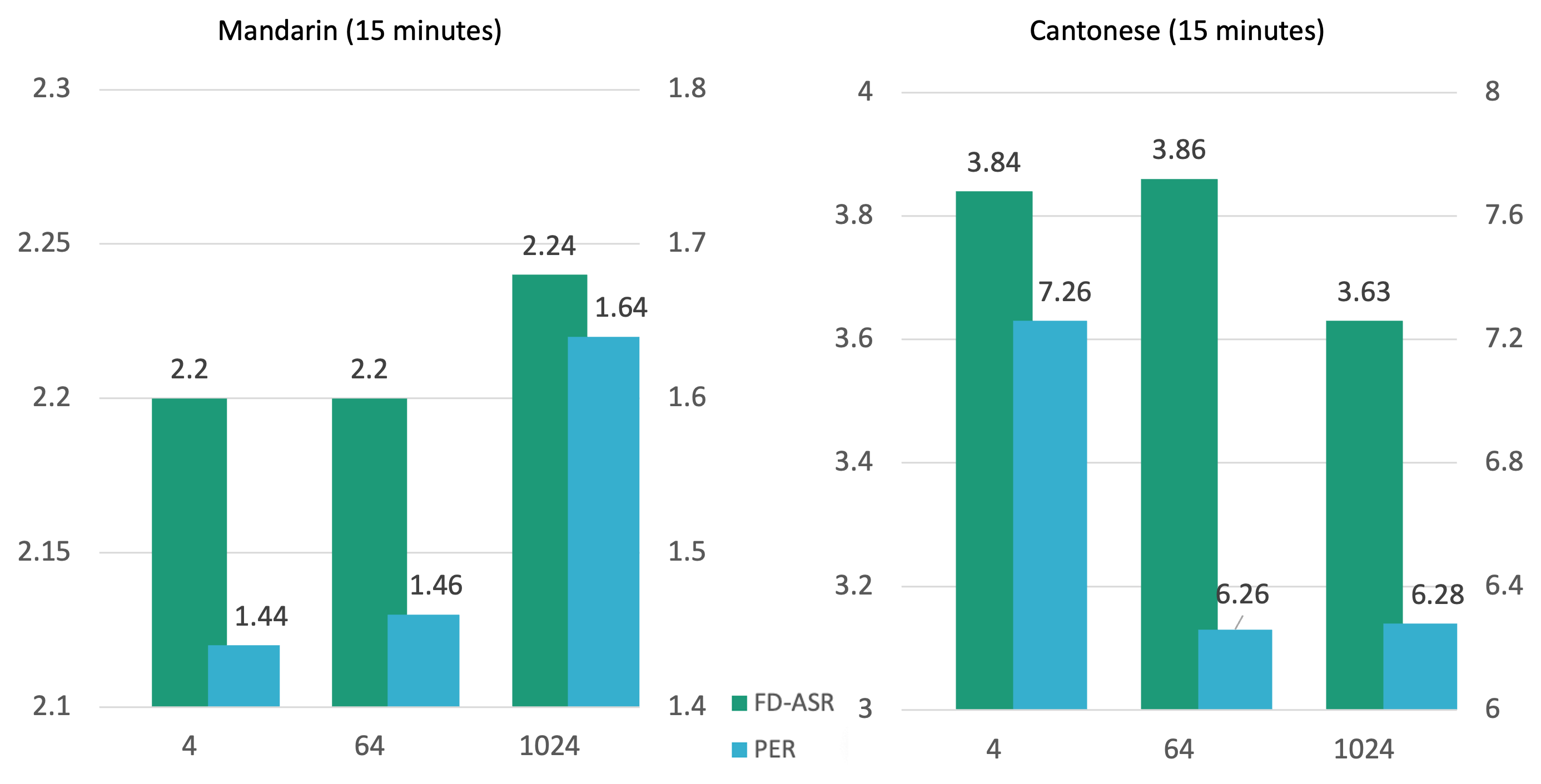}  
\caption{The impact of different codebook sizes to associate VQ-S3R learner in Mandarin and Cantonese low-resource TTS}
\label{fig:codebook}
\end{figure}

Besides, we also investigate the impact of the codebook size of the associate VQ-S3R learner on TTS. As shown in Fig. \ref{fig:codebook}, we train three VQ-VAE models with the codebook sizes of 4, 64, and 1024, then apply them for acoustic model training in Mandarin and Cantonese. The results in these two scenarios show opposite conclusions, that the Mandarin TTS system sharing the same language as the pre-training set prefers a smaller codebook size, while the Cantonese TTS system in a different language from the pre-training set prefers a larger codebook size. The highly-compact VQ sequence can abstract phonetic information in the Mandarin training set well, but also lacks generalization to represent cross-lingual speech. Hence, the larger codebook size benefits Cantonese TTS instead. In practice, we suggest training multiple learners with different codebook sizes, and selecting the suitable learner in TTS training in terms of the difference in language between the pre-training set and the supervised dataset.

\begin{figure}[htp]
    \includegraphics[width=8.8cm]{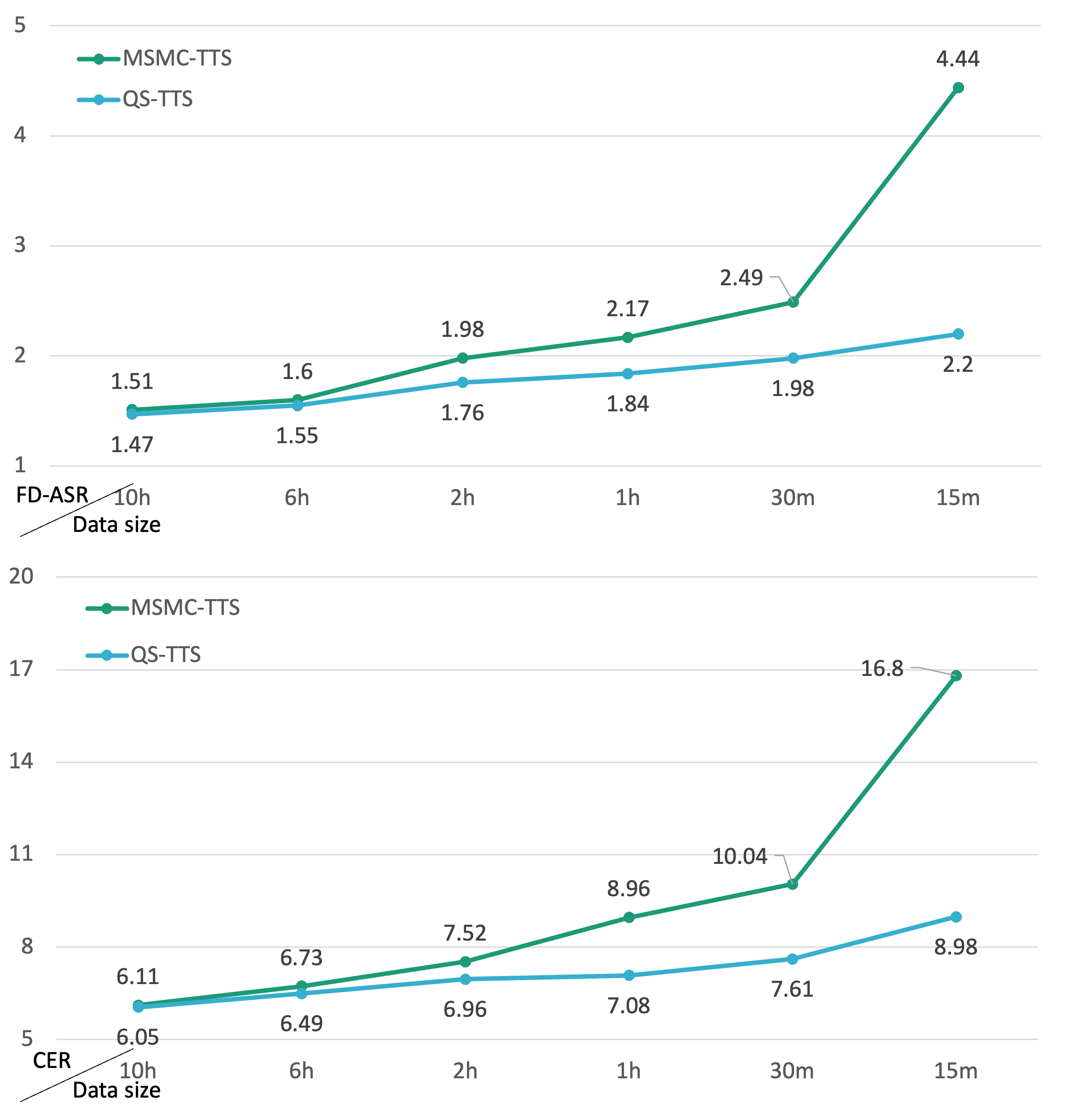}
    \caption{The FD-ASRs and PERs of MSMC-TTS and QS-TTS trained with different data sizes of CSMSC.}
    \label{fig:requirements}
\end{figure}

\subsection{Requirements for Supervised Data}

Finally, we investigate the supervised data requirements of QS-TTS by comparing MSMC-TTS and QS-TTS trained with different sizes of CSMSC, and drawing the line charts of FD-ASR and CER, as shown in Fig. \ref{fig:requirements}. First, the fully-supervised MSMC-TTS and semi-supervised QS-TTS achieve similar and good TTS performance with sufficient supervised data. Under the situation with 10-hour supervised data, the gap between these two systems is just 0.04 on FD-ASR and 0.06\% on CER. However, as data size decreases, the synthesis quality of MSMC-TTS degrades significantly, showing the rapidly increased FD-ASR and CER. In the low-resource scenario with only 15 minutes of supervised data, it achieves the FD-ASR of 4.44 and CER of 16.80\%, which are 193\% and 175\% higher than those in the system trained with 10-hour supervised data. Instead, QS-TTS shows a much-lower audio quality and intelligibility decay as supervised data size decreases. The QS-TTS trained with 15-minute supervised data achieves the FD-ASR of 2.2 and CER of 8.98\%, which are only 50\% and 48\% higher than QS-TTS trained with 10-hour supervised data. And the gap between these two approaches is also widened to 2.22 on FD-ASR and 7.82\% on CER, which is nearly a hundred times larger. Hence, under the low-resource scenario, MSMC-TTS has produced serious pronunciation issues, but QS-TTS can still keep good performance. This result strongly validates that the proposed semi-supervised TTS framework, QS-TTS, has a lower requirement for supervised data, and indicates its great potential in low-resource scenarios.

\section{Conclusion}
\label{sec:con}

This paper proposes QS-TTS, a novel semi-supervised TTS framework based on VQ-S3RL to effectively utilize more unlabeled speech audio to improve TTS quality while reducing its requirements for supervised data. The VQ-S3RL is conducted through two learners: The principal learner combines Multi-Stage Multi-Codebook (MSMC) VQ-GAN with contrastive S3RL to learn high-quality generative MSMC VQ-S3R, while decoding it to the high-quality audio; the associate learner further compresses the MSMCR into a highly-compact VQ representation via a VQ-VAE-based model. Then, TTS is implemented based on the MSMCR, and applied with VQ-S3R learners via transfer learning to achieve higher synthesis quality with lower supervised data requirements. This proposed framework can synthesize high-quality speech with lower supervised data requirements, significantly outperforming mainstream supervised and semi-supervised TTS approaches, especially in low-resource scenarios. Besides, the proposed VQ-S3RL also shows its effectiveness in providing better speech representations and pre-trained models for TTS by comparing with TTS systems with different speech representations and transfer learning methods. Finally, the slowly decayed performance of QS-TTS as supervised data decreases further validates its lower requirement for supervised data, and indicates its great potential in low-resource scenarios.

\bibliographystyle{unsrt}
\bibliography{refs}

\end{document}